\documentstyle[prl,aps,twoside,multicol]{revtex}

\begin{document}
\title{Creating Decoherence-Free Subspaces Using Strong and Fast Pulses}
\author{L.-A. Wu and D.A. Lidar}
\address{Chemical Physics Theory Group, University of Toronto, 80 St. George
Str., Toronto, Ontario M5S 3H6, Canada}
\date{\today }
\maketitle

\begin{abstract}
A decoherence-free subspace (DFS) isolates quantum information from
deleterious environmental interactions. We give explicit sequences of strong and fast (``bang-bang'', BB) pulses
that create the conditions allowing for the existence of
DFSs that support scalable, universal quantum computation. One such example is the creation of the
conditions for collective decoherence, wherein all system particles are
coupled in an identical manner to their environment. The BB pulses needed
for this are generated using only the Heisenberg exchange interaction. In
conjunction with previous results, this shows that Heisenberg exchange is
all by itself an enabler of universal fault tolerant quantum computation on
DFSs.
\end{abstract}


\begin{multicols}{2}

Since the discovery of quantum error correcting codes
(QECCs) \cite{Shor:95Steane:96a}, an arsenal of powerful methods has been
developed for overcoming the problem of decoherence that plagues quantum
computers (QCs). A QECC is a closed-loop procedure, that involves frequent
error identification via non-destructive measurements, and concommitant recovery steps. Alternatively,
decoherence-free subspaces (DFSs)
\cite{Zanardi:97c,Duan:98,Lidar:PRL98} and subsystems
\cite{Knill:99a},
and dynamical decoupling, or ``bang-bang'' (BB)
\cite{Viola:98,Vitali:99,Viola:99,Viola:99a,Zanardi:98bViola:00a,ByrdLidar:01},
are open-loop methods. A DFS is subspace of the system Hilbert space
which is isolated, by virtue of a dynamical symmetry, from the
system-bath interaction. The BB method is a close cousin of the
spin-echo effect. All decoherence-reduction
methods make assumptions about the system ($S$)-bath ($B$) coupling,
embodied in a Hamiltonian of the general form $H=H_{S}\otimes
I_{B}+H_{SB}+I_{S}\otimes H_{B}$. Here $I$ is an identity operator and $
H_{SB}$ is the system-bath interaction term, which can be expanded as a sum
over linear, bilinear, and higher order coupling terms: 
\begin{equation}
H_{SB}=\sum_{i} H_{i}+\sum_{i<j} H_{ij}+\ldots +\sum_{i_{1}<...<i_{p}}
H_{i_{1}i_{2}...i_{p}}.  \label{eq:HSB}
\end{equation}
Specializing to qubits, a typical assumption is $p=1$,
\begin{equation}
H_{i}=\vec{\sigma }_{i}\cdot\vec{B}_{i} =
\sum_{\alpha = x,y,z} \sigma_i^{\alpha} \otimes B_i^\alpha,
\end{equation}
where $\vec{\sigma }_{i}=(\sigma _{i}^{x},\sigma _{i}^{y},\sigma
_{i}^{z})\equiv (X_{i},Y_{i},Z_{i})$ are the Pauli matrices acting on the $
i^{{\rm th}}$ qubit, and $\vec{B}
_{i}=(B_{i}^{x},B_{i}^{y},B_{i}^{z})$ are arbitrary bath operators. This
includes the {\em independent errors model} (all $B_{i}^{\alpha }$
different, $\alpha =x,y,z$), and the {\em collective decoherence model} ($
B_{i}^{\alpha }\equiv B^{\alpha }$ $\forall i$), important, respectively for
the QECC and DFS methods. The bilinear terms are in general descibed by a
second-rank tensor ${\bf G} _{ij}$, so that $H_{ij}=\vec{\sigma }
_{i}\cdot {\bf G}_{ij}\cdot\vec{\sigma }_{j}$. One of the main
problems in applying the various decoherence-countering strategies is that,
typically, the conditions under which they apply are not wholly satisfied
experimentally \cite{Cory:98Kielpinski:01}. This problem is particularly
severe for the DFS method, since it demands a high degree of symmetry in the
system-bath interaction. Two main cases are known that admit {\em scalable} DFSs
(i.e., subspaces that occupy a finite fraction of the system Hilbert space):
collective decoherence \cite{Zanardi:97c,Duan:98,Lidar:PRL98} and
the model of ``multiple qubit errors'' (MQE) \cite{Lidar:00a}. Collective
decoherence, as defined above, assumes qubit-permutation-invariant
system-bath coupling. This may be satisfied at ultralow temperatures in
solid-state QC implementations, provided the dominant decoherence mechanism
is due to coupling to a long-wavelength reservoir, e.g., phonons \cite{Zanardi:98,Bacon:99a}. MQE assumes that the system terms appearing in $
H_{SB}$ generate an Abelian group under multiplication (referred to below as
the ``error group''). This is a somewhat artificial model that typically
imposes severe restrictions on $\vec{B}_{i}$ and ${\bf G}_{ij}$
(examples are given below). On the other hand, a two-dimensional DFS (encoding one
logical qubit) can be constructed using as few as 3 qubits under
collective decoherence conditions \cite{Knill:99a}, and unlike QECCs, requires no active intervention other than the
initial encoding and final decoding steps. One is thus faced with a rather
frustrating situation: the attractively simple DFS method imposes symmetry
demands that are likely to be perturbed in practice. Even though DFSs are
robust with respect to such symmetry-breaking perturbations \cite{Lidar:PRL98,Bacon:99}, and can be further stabilized by concatenation with
QECCs \cite{Lidar:PRL99}, it is highly desirable to be able to artificially
engineer conditions under which scalable DFSs may exist. Here we show how
this can be accomplished for both the collective decoherence and MQE models,
by combining the DFS encoding with the BB method. While
such ``environment engineering'' methods have been proposed before \cite{Zanardi:98bViola:00a,Poyatos:96}, we show here how this
can be accomplished for DFSs, assuming only physically reasonable resources.
In particular, we show that by using decoupling pulses that are generated
using only the isotropic Heisenberg exchange-interaction, one can transform
the general linear system-bath term $\sum_{i}H_{i}$ into a purely collective
decoherence term. Since the Heisenberg interaction
is by itself universal on the DFS-encoded qubits \cite{Bacon:99a,Kempe:00},
this result has direct implications for the promising QC proposals that make
use of the Heisenberg interaction to couple qubits \cite{Loss:98Levy:01a,Kane:98Khitun:01,Vrijen:00}, and in which the use of
single-qubit operations is preferably avoided \cite{DiVincenzo:00a,LidarWu:01,Kempe:01}.

{\it Dynamical decoupling and DFS}.--- Let us start by briefly reviewing the
decoupling technique, as it pertains to our problem (for a thorough review
see, e.g., \cite{Zanardi:98bViola:00a}). Decoupling relies on the ability to
apply strong and fast (BB) pulses \cite{Viola:98}, in a manner which effectively averages $H_{SB}$ to
zero. Since the pulses are strong one
ignores the evolution under $H_{SB}$ while the pulses are on, and
since the pulses are fast one makes the short-time approximation,
i.e., $\exp[(A+B)t] \approx \exp(A t)\exp(B t)$ for $[A,B]\neq 0$. Systematic
corrections are known \cite{Viola:98}.
The simplest example of eliminating an undesired unitary evolution $U=\exp
(-itH_{SB})$ is the {\em parity-kick sequence}
\cite{Viola:98,Vitali:99}. Suppose we have at our disposal a fully
controllable interaction generating a gate $R$ such that ``$R$ {\em
conjugates} $U$'': $R^{\dagger }UR=U^{\dagger }$. Then the sequence
$UR^{\dagger }UR=I$ serves to eliminate $U$.

Now, turning on the single-qubit Hamiltonian $\epsilon_i^x X_i$ for a time $
t=\pi/2\epsilon_i^x$ generates the single-qubit gate $
X_{i}=i\exp (-i\frac{\pi }{2}X_{i})$. Each term in $H_{SB}$ either commutes
or anti-commutes with $X_{i}$ since each term contains at least one factor
of $ X_{i},Y_{i},Z_{i}$. {\em We call a term} $A$ {\em ``even'' with respect
to} $B$ {\em if} $[A,B]=0$, {\em ``odd'' if} $\{A,B\}=0$. If a term $A$ in
$H_{SB}$ is odd with respect
to (wrt) $X_{i}$ then the evolution
under it will be conjugated by the gate $X_{i}$: $X_{i}\exp (-iA\tau)X_{i}=\exp (iA\tau )$. This allows for selectively removing this term using
the parity-kick cycle, which we write as:\ $[\tau ,X_{i},\tau ,X_{i}]$.
Reading from right to left, this notation means: apply $X_{i}$ pulse, free
evolution for time $\tau $, repeat. Since every system factor in $H_{SB}$
contains a single-qubit operator, it follows that we can selectively keep or
remove each term in $H_{SB}$ by using the parity-kick cycle. Note, however,
that in general we have to use a short-time approximation since $
[U,R^{\dagger }UR]\neq 0$ \cite{Vitali:99}. Further, without additional
symmetry assumptions restricting $p$ of Eq.~(\ref{eq:HSB}), this procedure, if used to eliminate 
{\em all errors}, requires a number of pulses that is exponential in $N$ 
\cite{Viola:99}. The reason is that without symmetry we will need at least
two non-commuting single-qubit operators per qubit (e.g., $X_{i}$,
$Y_{i}$).

The DFS method uses a very different idea for overcoming
decoherence \cite{Zanardi:97c,Duan:98,Lidar:PRL98}. Suppose that there exist states $\{ |\psi_i \rangle \}$
that are degenerate under the action of all system operators
$S_\alpha$ in the interaction Hamiltonian $H_{SB} = \sum_\alpha
S_\alpha \otimes B_\alpha$, i.e.: $S_\alpha |\psi_i \rangle = c_\alpha
|\psi_i \rangle$ $\forall \alpha,i$, where $c_\alpha$ are
constants (more general conditions can be found \cite{Knill:99a}). Such a collection of states forms a subspace that acquires
only an overall phase under the action of $H_{SB}$, and is therefore
decoherence-free. The requisite degeneracy arises from a symmetry in
$H_{SB}$, such as collective decoherence.

{\it Symmetrization}.--- We now turn to showing how decoupling may be used
to create the conditions for DFSs. General group-theoretic arguments for
using BB pulses for ``symmetrizing'' system-bath interactions, thus creating
DFS conditions, were given in \cite{Zanardi:98bViola:00a}. However, these
proposals did not consider the MQE model and did not give an explicit
Hamiltonian realization for the collective decoherence model. Here we
specialize to the MQE and collective decoherence models, and give explicit
pulse sequences that respect the constraints imposed by physically available
resources.

{\it Generation of the MQE model}.--- The MQE model assumes that the system
operators appearing in $H_{SB}$ form an Abelian group ${\cal G}$ under
multiplication \cite{Lidar:00a}. E.g., $H_{SB}=\sum_{i=1}^{N-1}Z_{i}Z_{i+1}
\otimes B_{i}$, or $H_{SB}=\sum_{i=1}^{N/2}\sum_{\alpha =x,y,z}\sigma
_{2i-1}^{\alpha }\sigma _{2i}^{\alpha }\otimes B_{i}^{\alpha }$. The
dimension of the DFS supported by ${\cal G}$ is $2^{N}/|{\cal G}|$, where $| 
{\cal G}|$ is the order of ${\cal G}$, which also counts the number of
independent errors the DFS is immune to \cite{Lidar:00a}. An Abelian group
with $M$ generators has order $2^{M}$. Universal, fault-tolerant quantum
computation can be performed on the MQE class of DFSs using the method of 
\cite{Lidar:00b}. Briefly, this method involves a hybrid DFS-QECC approach,
wherein logic gates acting on DFS states are supplemented with
fault-tolerant error detection and recovery. This active QECC intervention
is needed since, unlike the collective decoherence case treated below, in
the MQE case logic gates take encoded states on a trajectory that begins
inside the DFS, leaves it, and then returns (as is also the case for
computation using QECCs). Let us now show how to generate
the MQE conditions starting from a $p=2$ system-bath Hamiltonian.

We assume that {\em single-qubit gates are available}. In this case, it has
been shown that the linear term, $\sum_{i}H_{i}$, can be eliminated using $4$
pulses, each acting simultaneously on all qubits \cite{Viola:99,Viola:99a}.
We reproduce this result and show further how single-qubit gates can
efficiently decouple bilinear Hamiltonians $H_{ij}$ with nearest-neighbor
interactions. Let $H_{{\rm nn}}=\sum_{i=1}^{N}H_{i}+H_{i,i+1}$ and $U_{{\rm 
nn}}=\exp (-i\tau H_{{\rm nn}})$. Define collective rotation operators 
\begin{equation}
R=R_{1}R_{2}\cdots R_{N},\quad R_{O}=R_{1}R_{3}\cdots R_{N/2-1} 
\label{eq:R}
\end{equation}
where $N$ is even and $R$ can be $X$, $Y$ or $Z$. First note that $U_{{\rm 
nn }}^{^{\prime }}=U_{{\rm nn}}(XU_{{\rm nn}}X)$ leaves only those linear
terms containing $X_{i}$, and all bilinear terms of the form $\sigma
_{i}^{\alpha }\sigma _{j}^{\alpha }$, $Y_{i}Z_{j}$. Let us apply $Z$ to the
outcome, i.e., $U_{{\rm nn}}^{^{\prime \prime }}=U_{{\rm nn}}^{^{\prime
}}(ZU_{{\rm nn}}^{^{\prime }}Z)$. This eliminates all linear terms in $4$
pulses: $[\tau ,Y,\tau ,X,\tau ,Y,\tau ,X]$ (where we have used $Y=-iXZ$).
It also eliminates all $Y_{i}Z_{j}\otimes B_{ij}^{yz}$, leaving just $
\sum_{j}\sigma _{j}^{\alpha }\sigma _{j+1}^{\alpha }\otimes
B_{j,j+1}^{\alpha \alpha }$ ($ \alpha =x,y,z$). This we can rewrite as $
\sum_{j}\sigma _{j}^{\alpha }\sigma _{j+1}^{\alpha }\otimes
B_{j,j+1}^{\alpha \alpha }=\sum_{j={\rm odd}}$ $\vec{\sigma }
_{j}\cdot\vec{B^{\prime }}_{j}$, i.e., the even-numbered qubits
act as baths for the odd-numbered qubits. Now let $ U_{{\rm nn}}^{^{\prime
\prime \prime }}=U_{{\rm nn}}^{^{\prime \prime }}(X_{O}U_{{\rm nn}
}^{^{\prime \prime }}X_{O})$, which requires $8$ pulses. At this point we
are left only with errors of the form $X_{j}X_{j+1}$. These generate an
Abelian group denoted $Q_{2X}$\ in \cite{Lidar:00b}. Since $Q_{2X}$ has $N-1$
generators its order is $|Q_{2X}|=2^{N-1}$, so that the DFS is $
2^{N}/2^{N-1}=2$ dimensional, i.e., supports a single encoded qubit, which
of course is not scalable. A larger number of encoded qubits can be
supported by reducing the dimension of the error group. This, in turn,
requires a few more pulses. E.g., consider applying the BB pulse $
Z_{3}Z_{4}Z_{7}Z_{8}\cdots $ to $U_{{\rm nn}}^{^{\prime \prime \prime }}$.
This eliminates $X_{2}X_{3},X_{4}X_{5},X_{6}X_{7},...$. What is left, after $
16$ pulses, is the error group generated by $\{X_{2j-1}X_{2j}\}_{j=1}^{N/2}$, denoted $Q_{X}$ in \cite{Lidar:00a}. It has order $|Q_{X}|=2^{N/2} $, thus
supporting a $2^{N}/2^{N/2}=2^{N/2}$ dimensional DFS. This DFS encodes $N/2$
qubits, so it is scalable. The methods of \cite{Lidar:00b} now apply for the
purpose of fault-tolerant universal quantum computation using the hybrid
DFS-QECC method.

Note that we can also go further and eliminate all second order coupling
terms: $U_{{\rm nn}}^{^{\prime \prime \prime }}(Z_{O}U_{{\rm nn}}^{^{\prime
\prime \prime }}Z_{O})=I$ and also uses a total of $16$ collective pulses.
If there is a next-nearest-neighbor system-bath interaction, it can
similarly be removed using the collective operator $
R_{OO}=R_{1}R_{5}R_{9}R_{13}\cdots $, etc. for longer-range interactions. These manipulations
will leave higher order MQE models; which option to choose will depend on
which pulse sequences are most easily implementable. It should be clear that
this method of generating MQE models is quite general: given a system-bath
Hamiltonian, one can design a set of BB pulses that will transform this
Hamiltonian into a desired Abelian error group. The number of pulses will
scale with the system-bath coupling order $p$ [Eq.~(\ref{eq:HSB})] and the interaction range $r$ ($r=1$ for
nearest-neighbors, etc.), and we have given explicit
sequences for the case $p,r\leq 2$. We note that this combination of
decoupling with a hybrid DFS-QECC strategy is, as far as we know, the first
time that {\em all three methods for combatting decoherence have been used
together}.

As a final comment on generating MQE models, we note that the analysis above
applies also to the case where it may be preferable to control two-qubit ``product''
Hamiltonians of the form $X_{i}X_{j}$ and $ Y_{i}Y_{j}$. Similar to the case
of controllable single-qubit gates, we now have the gates $X_{i}X_{j}=i\exp
(-i\pi X_{i}X_{j}/2)$ and $ Y_{i}Y_{j}=i\exp (-i\pi Y_{i}Y_{j}/2)$. Such
gates could be implemented naturally, e.g., in certain superconducting QC
implementations \cite{Spiller:00}. It is simple to check that the product Hamiltonians can be used to
decouple any linear system-bath Hamiltonian, and any bilinear term other
than $\sigma _{i}^{\alpha }\sigma _{j}^{\alpha }$. In fact, we can construct
the $R$ gates [Eq.~(\ref{eq:R})] by simply turning on all nearest neighbor gates $\sigma
_{j}^{\alpha }\sigma _{j+1}^{\alpha }$, and the $R_{O}$ gates by
simultaneously turning on all next-nearest neighbor gates $\sigma
_{j}^{\alpha }\sigma _{j+2}^{\alpha }$. Therefore the methods used above for
the case of single-qubit gates apply directly.

{\it Generating collective decoherence}.--- We now turn to 
collective decoherence. To do so, we consider the
important case of a controllable Heisenberg exchange Hamiltonian $J_{ij}
\vec{\sigma }_{i}\cdot\vec{\sigma }_{j}$, crucial for
the operation of, e.g., quantum dot QCs \cite{Loss:98Levy:01a} and donor atom
nuclear \cite{Kane:98Khitun:01} or electron \cite{Vrijen:00} spin QCs. We
will show that {\em using a few collective pulses generated by the
Heisenberg interaction alone, we can symmetrize any linear system-bath
interaction }$\sum_{i}\vec{\sigma }_{i}\cdot\vec{B}
_{i}$, {\em such that only a block-collective component remains}.
This collective decoherence can then be avoided by encoding into a 4-qubit
DFS \cite{Zanardi:97c}, or a 3-qubit DFS/noiseless subsystem (NS)
\cite{Knill:99a}. Let
\[
O_{ij}\equiv \exp (-i\pi\vec{\sigma }_{i}\cdot\vec{
\sigma }_{j}/4)
\]
A simple calculation shows that $O_{ij}$ is a {\sc SWAP} gate for the Pauli
matrices: $O_{ij}^{\dagger }\vec{\sigma }_{i}O_{ij}=
\vec{\sigma }_{j}$. From this follows $O_{ij}^{\dagger }\left( 
\vec{\sigma }_{i}\pm\vec{\sigma }_{j}\right)
O_{ij}=\pm \left(\vec{\sigma }_{i}\pm\vec{\sigma }
_{j}\right) $, i.e., all {\em differences} $\vec{\sigma }_{i}-
\vec{\sigma }_{j}$ are odd wrt $O_{ij}$, and hence can be
eliminated$.$ Let us then rewrite $\sum_{i=1}^{N}\vec{\sigma }
_{i}\cdot\vec{B}_{i}=\sum_{\beta =\pm }\sum_{j=1}^{N/2}\left( 
\vec{\sigma }_{2j}+\beta\vec{\sigma }_{2j-1}\right)
\cdot \vec{B}_{2j}^{\beta }$, where $\vec{B}_{2j}^{\pm }\equiv (\vec{B}
_{2j}\pm \vec{B}_{2j-1})/2$. To eliminate the nearest-neighbor differences $
\left(\vec{\sigma }_{2j}-\vec{\sigma }_{2j-1}\right) $
we can use the collective BB pulse $O=\bigotimes_{j=1}^{N/2}O_{2j-1,2j}$.
This leaves only the (block-)collective decoherence term $
\sum_{j=1}^{N/2}\left(\vec{\sigma }_{2j}+\vec{\sigma }
_{2j-1}\right) \cdot \vec{B}_{j}^{+}$, which in turn we can rewrite as $
\sum_{\beta =\pm }\sum_{j=1}^{N/2}\left(\vec{\sigma }_{2j+2}+
\vec{\sigma }_{2j+1}+\beta \left(\vec{\sigma }_{2j}+
\vec{\sigma }_{2j-1}\right) \right) \cdot \vec{B}_{2j}^{+,\beta }$, where $\vec{B}_{2j}^{+,\pm }\equiv (\vec{B}_{2j+2}^{+}\pm \vec{B}
_{2j}^{+})/2$. We can now eliminate the next-nearest neighbor differences $
\left(\vec{\sigma }_{2j+2}-\vec{\sigma }_{2j}\right) $
and $\left(\vec{\sigma }_{2j+1}-\vec{\sigma }
_{2j-1}\right) $ using a second collective pulse $O_{O}=
\bigotimes_{j=1}^{N/2-1}O_{2j-1,2j+1}O_{2j,2j+2}$. At this point we are left
just with the collective decoherence terms on blocks of $4$ qubits, and the
encoding into the 4-qubit DFS \cite{Zanardi:97c} becomes relevant. This
scheme uses a total of $6$ collective BB pulses: $[O_{O},O,\tau ,O^{\dagger},\tau ,O_{O}^{\dagger },O,\tau ,O^{\dagger }]$. The apparent drawback of
needing next-nearest neighbor interactions (in the $O_{O}$ pulse) can be
avoided by swapping local gates, e.g.: $O_{i,i+2}=$ $O_{i+1,i+2}^{\dagger
}O_{i,i+1}O_{i+1,i+2}$, at the expense of more pulses. Note that in a 2D
hexagonal arrangement $i,i+1$ and $i,i+2$ can all be nearest neighbors.

We can also symmetrize into blocks of $3$, which can be used for the $3$-qubit DFS/NS \cite{Knill:99a}. Let us rewrite $H_{3}=\sum_{i=1}^{3} 
\vec{\sigma }_{i}\cdot\vec{B}_{i}=\left( 
\vec{\sigma }_{1}+\vec{\sigma }_{2}+\vec{
\sigma }_{3}\right)\cdot \vec{A}^{+} +\left(\vec{\sigma }_{2}-\vec{
\sigma }_{1}\right)\cdot \vec{A}^{-} +\vec{\sigma }_{3} \cdot \vec{C}$, where $\vec{A}^{\pm}\equiv
(\vec{B}_{2} \pm \vec{B}_{1})/2$ and $\vec{C} \equiv \vec{B}_{3}- \vec{A}^{+}$. We can eliminate $\vec{
\sigma }_{2}-\vec{\sigma }_{1}$ using $O_{12}$: $
U_{1}(\tau)=\exp (-iH_{3}\tau)O_{12}^{\dagger }\exp (-iH_{3}\tau)O_{12}=\exp 
\left[ -2i\tau( \left(\vec{\sigma }_{1}+\vec{\sigma }
_{2}+\vec{ \sigma }_{3}\right)\cdot \vec{A}^{+} +\vec{\sigma
}_{3}\cdot \vec{C} \right] $. Next consider 
\begin{eqnarray}
U_{2}(\tau) &=& U_{1}(\tau/2)O_{23}^{\dagger }U_{1}(\tau)O_{23}  \nonumber \\
&=& e^{ -i\tau\left[ 3\left(\vec{\sigma }_{1}+\vec{
\sigma }_{2}+\vec{\sigma }_{3}\right)\cdot \vec{A}^{+}+\left( 2\vec{
\sigma }_{2}+\vec{\sigma }_{3}\right)\cdot \vec{C} \right] }.  \nonumber
\end{eqnarray}
Finally,
\[
U_{2}(\tau)O_{12}^{\dagger }U_{2}(\tau)O_{12}= e^{ -i\tau \left( 
\vec{\sigma }_{1}+\vec{\sigma }_{2}+\vec{
\sigma }_{3}\right)\cdot \left( \vec{B}_1 + \vec{B}_2 + \vec{B}_3 \right) }, 
\]
leaving only the collective component. This scheme uses a total of $14$
collective BB pulses.

To quantum compute universally on DFSs it is necessary to couple blocks of
DFS qubits, in order to implement a controlled-logic gate
\cite{Bacon:99a,Kempe:00}. An extra symmetrization step is thus required,
creating collective decoherence conditions over blocks of $6$ or $8$ qubits.
This is a simple extension of the procedure above. E.g., for two coupled $4$-qubit DFS blocks, we need collective pulses of the form $ O_{OO}=
\bigotimes_{j=1}^{N/2-3}O_{2j-1,2j+3}O_{2j,2j+4}O_{2j+1,2j+5}O_{2j+2,2j+6}$.
By swapping local gates we can again avoid direct control over long-range
interactions. The corresponding increase in the number of gates may well be
a worthwhile tradeoff. Similar pulse sequences can be found for creating
block-collective decoherence conditions over $6$ qubits, for computation
using the $3$-qubit NS.

{\it Discussion and conclusions}.--- The prospect of decoherence-free
quantum computation is very attractive, but so far ideas for obtaining the
conditions enabling the existence of scalable decoherence-free subspaces
(DFSs) focused mostly on lowering the temperature and neglecting other
sources of decoherence \cite{Zanardi:98,Bacon:99a}. The exception are
previous existential results showing how decoupling methods can be used for
symmetrization of system-bath interactions \cite{Zanardi:98bViola:00a}. In
this work we showed explicitly how conditions for the two most important
examples of DFSs (the models of collective decoherence
\cite{Zanardi:97c,Duan:98,Lidar:PRL98} and multiple qubit
errors \cite{Lidar:00a}, MQE) can be actively generated using
symmetrizing cycles of 
fast and strong decoupling (``bang-bang'', BB) pulses. In the MQE case a
cycle of $16$ pulses suffices to symmetrize a system-bath Hamiltonian with
arbitrary linear and nearest-neighbor bilinear couplings, such that
conditions enabling the existence of a scalable DFS are established. This
result is applicable for quantum computer (QC) proposals where single-qubit
gates are easily tunable. In this case a hybrid DFS-active
quantum error correction scheme developed in \cite{Lidar:00b}, can be used
for universal, fault-tolerant quantum computation.

In the case of collective decoherence a very attractive general picture is
emerging, from a combination of previous studies and this work. The
collective decoherence model was first proposed as an example allowing the
existence of DFSs with the property of a scalable encoding \cite{Zanardi:97c,Duan:98,Lidar:PRL98}. It was later realized that
universal quantum computation is possible on these DFSs using the Heisenberg
exchange interaction alone \cite{Bacon:99a,Kempe:00}. Our present result
shows how {\em collective decoherence conditions can be actively created
using a few pulses generated using only Heisenberg exchange}. Since our
method relies on BB pulses some degree of leakage out of the DFS (due to
imperfect symmetrization) is inevitable. Fortunately, such leakage
errors can also be reduced using Heisenberg-generated BB pulses \cite{WuByrdLidar:02}, or detected with a circuit that utilizes, again, only Heisenberg
exchange \cite{Kempe:01}.
The combination of all these results shows that {\em Heisenberg exchange is
all by itself an enabler of universal fault tolerant quantum computation on
decoherence-free subspaces}. This has potentially important applications for
those solid-state proposals of QCs where Heisenberg exchange is the natural
qubit-qubit coupling mechanism \cite{Loss:98Levy:01a,Kane:98Khitun:01,Vrijen:00}. The combination of the decoupling method with encoding methods developed
in the quest to protect fragile quantum information thus seems to be a
promising route towards robust implementations of QCs.

{\it Acknowledgements.---} The present study was sponsored by the
DARPA-QuIST program (managed by AFOSR under agreement
No. F49620-01-1-0468), and by PRO (to D.A.L.).

\end{multicols}

\end{document}